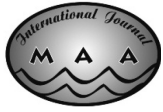

**RESEARCH ARTICLE**

# THE DEVELOPMENT OF A UTOPIAN CITY? COMPARING LAND- AND SKYSCAPES IN SANTA CRUZ DE TENERIFE AND SAN CRISTOBAL DE LA LAGUNA

**Alejandro Gangui*[1,2] and Juan Antonio Belmonte[3]**

[1]*Universidad de Buenos Aires, Facultad de Ciencias Exactas y Naturales, Argentina.*
[2]*CONICET - Universidad de Buenos Aires, Instituto de Astronomía y Física del Espacio (IAFE), Argentina.*
[3]*Instituto de Astrofísica de Canarias and Universidad de La Laguna, Tenerife, Spain.*

**ABSTRACT**

We discuss the peculiar planning of the city of San Cristóbal de La Laguna, in the Canary Island of Tenerife (Spain), when compared to the nearby and essentially contemporary Santa Cruz de Tenerife, which served as a maritime port of the former city. For this we review our previous study of the exact spatial orientation of twenty-one historic Christian churches currently existing in the old part of La Laguna, which we compare with the analysis of six similar buildings located in Santa Cruz, and presented here for the first time. In both cities, we take the spatial orientation of historic churches as good indicators of the original layout of the respective urban lattices. Although we find a clear orientation pattern for La Laguna, which singles out an absolute-value astronomical declination slightly below 20°, pointing to a preferred date close to the July 25th feast-day of San Cristóbal de Licia, in the case of Santa Cruz this trend is not followed. On the contrary, the pattern we find for Santa Cruz, within the uncertainties due to the low statistics, and apart from one equinoctial and one solstitial oriented churches, is consistent with an orographic orientation within the canonic limits of sunrise. This result highlights the uniqueness of the city of La Laguna, and supports the idea suggesting its deliberate planning in the early 16th century.

## 1. INTRODUCTION

The city of San Cristóbal de La Laguna has an exceptional value due to the original conception of its plan. It is an urban system in a grid, outlined by straight streets that form squares, its layout being the first case of an unfortified colonial city with a regular plan in the overseas European expansion (UNESCO, 1999).

Common wisdom surrounding La Laguna suggests that the city, which was founded in 1496, would have been laid out according to utopian assumptions inspired by Greek philosophical principles (Navarro Segura, 1999). The old city, whose historical center was declared a World Heritage Site by UNESCO in 1999, would hence be the materialization of a especially conceived geometric plan, even including the golden proportion (Herráiz Sánchez, 2007), with deep roots in Plato's design for the ideal city of Magnesia, a main topic of the philosopher's last and unfinished work "Laws".

Recently, by analysing the exact spatial orientation of the more than twenty historic Christian churches and chapels currently existing in La Laguna, we proposed a way to falsify those far-fetched hypotheses (Gangui and Belmonte, 2018). This was part of a larger Canarian church orientation and planimetry project (Gangui et al., 2014). We were able to determine that a definite orientation pattern was indeed followed in La Laguna, but it was probably much more simple than imagined.

By considering the group of existing historic Christian churches as a good indicator of the original layout of the urban lattice in La Laguna, we found some emblematic churches oriented in the canonical way (McCluskey, 2015), namely towards the equinoxes, and a couple more aligned with the summer solstice, a characteristic time-mark of the island aboriginal population before the conquest (Belmonte, 2015). But most importantly, we were able to show that the general pattern singles out an orientation compatible with the feast day of Saint Christopher of Lycia (San Cristóbal de Licia), the saint to whom the city (i.e. San Cristóbal de La Laguna) was originally dedicated (see Figure 1).

Let us note that the declination histogram presented in Figure 1 is given in absolute value, which allows the possibility of both the rising and setting Sun as orientation targets. The figure shows the astronomical declination versus the normalised relative frequency. The continuous (dashed) vertical line represents the absolute value declination corresponding to the extreme position of the Sun (Moon) at the solstice (at major lunastice). There appears one outstanding peak dominating the chart which, although it is hard to identify with a single preferred date due to the spread in the data, as reflected in the width of the curve, as we mentioned, it might be associated with a date close to the feast-day of Saint Christopher of Lycia.

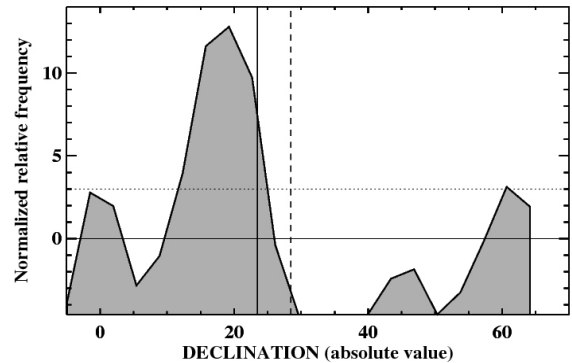

**Figure 1.** Absolute value declination histogram for the churches of La Laguna. Only one prominent peak slightly below c. 20° is found above the 3σ level (dotted horizontal line) and may be associated with the Catholic church feast-day of Saint Christopher of Lycia on July 25th (or 10th, both dates being relevant), in the Julian calendar, closely corresponding to the Sun's declination on that day at the time of the founding of the city. In addition to the main peak, there appear two barely statistically significant minor peaks. One is found around the equinox, pointing to a canonical orientation pattern, while the other -corresponding to a declination around 60°-, might be associated with an accumulation peak due to orientations near the meridian line. Adapted from Gangui and Belmonte (2018).

In order to test whether this result is exclusive of La Laguna, in this paper we extend our methods and measurements to another city of the island, Santa Cruz de Tenerife, nearly as old as La Laguna, and which is relatively close to it, as it fulfills the role of being its sea port. As we will see the pattern of orientations we find for Santa Cruz is qualitatively different, and this lends support to the hypothesis that there might have being a deliberate plan underlying the foundation of La Laguna starting around the year 1500.

## 2. METHODS AND DISCUSSION

In our recent work in La Laguna we found two important buildings (apparently) oriented close to the northern hemisphere winter solstice rising Sun, namely the churches of San Agustín and of Nuestra Señora de La Concepción, with azimuths 114½° and 115½° respectively (Gangui and Belmonte, 2018). We argued then that the most probable explanation for this was not given by their orientation towards the eastern horizon, as the winter solstice rising Sun was a very rare target in the Iberian Christian world (González-García and Belmonte, 2015), but towards the setting Sun during the opposite (summer) solstice (see Figure 2). This would be clearly a reasonable political and social solution for the Church to indulge the original Guanche population who inhabited those lands, and also the first colonists coming to Tenerife from the nearby island of Gran Canaria, for whom the summer solstice, contemporary to the time of harvest, was a much more relevant temporal milestone (Belmonte, 2015).

In line with this, we considered that builders of the churches in La Laguna, for issues related to orography or for other reasons, might have allowed actual orientations for both the rising and the setting Sun directions. We therefore included measurements of the buildings' axes in the opposite direction to the canonical one, namely towards the narthex of the churches (Gangui and Belmonte, 2018). We think it is important to do the same kind of analysis while employing the new measurements for the churches of Santa Cruz de Tenerife, as we will see in the data presented below, where we allow the possibility of both the rising and setting Sun as orientation targets.

Table 1 shows the new data resulting from our fieldwork in Santa Cruz. The identification of the churches is presented along with their geographical location in coordinates and orientation (archaeoastronomical data): the measured azimuth (rounded to ½° approximation) and the angular height of the point of the horizon towards which the altar or the narthex of the church is facing, as well as the derived computed declination corresponding to the central point of the solar disc. The measured height of the horizon was appropriately corrected for atmospheric refraction (Schaefer, 1993). When the horizon was blocked, we employed the digital elevation model based on the Shuttle Radar Topographic Mission (SRTM) available at HeyWhatsThat (Kosowsky, 2017), which gives angular heights within a ½° approximation.

We obtained our measurements using a tandem instrument Suunto 360PC/360R which incorporates a clinometer and a compass with a precision of ½° and also by analysing the surroundings (landscape) of each of the buildings. We then corrected the azimuth data according to the local magnetic declination (Natural Resources Canada, 2015), getting values always close to 5°07' W for different sites of Santa Cruz. Our data is the result of several on-site measurements with a single instrument, taking the axes of the churches, from the back of the buildings towards the altars, as our main guide. In some cases, although not in all, as many churches are surrounded by modern buildings, we could verify that the lateral walls were parallel to their axes. We estimate the error of our measurements to be around ±¾° for the resulting declination, by performing a simple propagation of errors. However, given the nature of the measurements, some of them obtained in the middle of the town and surrounded by asphalt, metal, and wires, we prefer to be more conservative and take our estimated error to be ±1° (upper bound).

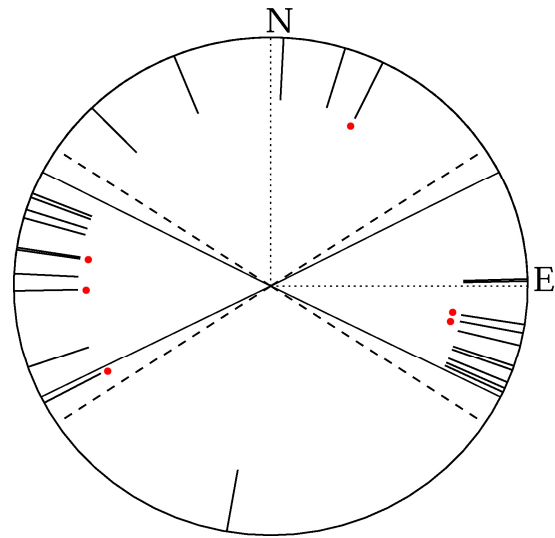

**Figure 2.** Orientation diagram of the churches and chapels of both cities: San Cristóbal de La Laguna and Santa Cruz de Tenerife (these last ones are signalled with a "dot" interior to the corresponding bar). These orientations correspond to the main axes of the measured churches, considering the direction from the front door towards the altar. The majority of the monuments follow the canonical orientation pattern within the solar range, while a few small private chapels of La Laguna (and Nuestra Señora de Regla in Santa Cruz) concentrate in the northern quadrant. The diagonal lines on the graph indicate, in the eastern quadrant, the extreme values of the corresponding azimuth for the Sun (azimuths of 62°.7 and 116°.6 –continuous lines– which are equivalent to the northern hemisphere summer and winter solstices, respectively) and for the Moon (azimuths of 56°.6 and 123°.6 –dotted lines– which are equivalent to the position of the major lunistices or lunar standstills).

From the data included in the Table we note that just two of the churches have a standard orientation. In the first place, San Sebastián, which has a very small declination value and is therefore presumably oriented towards the equinox. Secondly, Nuestra Señora del Pilar which is, to a very good approximation, solstitially oriented. This last characteristic is not difficult to understand, as the church is located in what once were the outskirts of the city, along the way leading to Anaga Peninsula, an area where the survival of earlier aborigine customs could have been greater than in the nucleus of the city closer to the harbour (Belmonte and Sanz de Lara, 2001).

**Table 1.** Orientations for the historic churches of Santa Cruz de Tenerife (see Figure 3): for each of them we show the identification (name and most likely date of construction), the geographical latitude and longitude (L and l), the astronomical azimuth (a) taken along the axis of the building towards the altar or towards the narthex (rounded to ½° approximation), the angular height of the horizon (h) in that direction (including the correction due to atmospheric refraction, and also rounded to ½° approximation) and the corresponding resultant declination (δ). Angular heights of the horizon for directions blocked by modern buildings within the city were obtained from numerical terrain models. The final column for the declination shows a combination of both δ (altar) and δ (narthex) in order to emphasize its absolute value, as explained in the text. Finally, the Orientation column is computed by estimating the dates (either Julian or Gregorian taking into account the documented year of the construction of each church) when the final declination of the Sun is the one indicated.

| NAME (DATE) | L (°/′) North | l (°/′) West | a (°) (altar) | h (°) (altar) | δ (°) (altar) | a (°) (narthex) | h (°) (narthex) | δ (°) (narthex) | δ (°) (final) | Patron saint date / Orientation |
|---|---|---|---|---|---|---|---|---|---|---|
| San Sebastián (XVI c.) | 28.4642 | 16.2550 | 269 | 5 | 1½ | 89 | 0 | 1 | 1 | 20 Jan / 14 Mar – 10 Sep |
| Ntra Sra de Regla (XVII c.) | 28.4569 | 16.2523 | 26 | 4½ | 56 | 206 | 1½ | -51 | 56 | 8 Sep / -------------------- |
| San Telmo (XVI c.) | 28.4626 | 16.2489 | 101 | 0 | -9½ | 281 | 5 | 12 | 12 | 14 Apr / 12 Apr – 12 Aug |
| N.S. de La Concepción (XVI c.) | 28.4645 | 16.2493 | 99 | 0 | -8 | 279 | 5 | 10 | 10 | 8 Dec / 6 Apr – 18 Aug |
| San Francisco (XVII c.) | 28.4679 | 16.2493 | 279 | 4½ | 10 | 99 | 0 | -8 | 10 | 4 Oct / 15 Apr – 27 Aug |
| Ntra Sra del Pilar (XVIII c.) | 28.4693 | 16.2519 | 242 | 4½ | -22 | 62 | 0 | 24½ | 24½ | 12 Oct / 21 Jun |

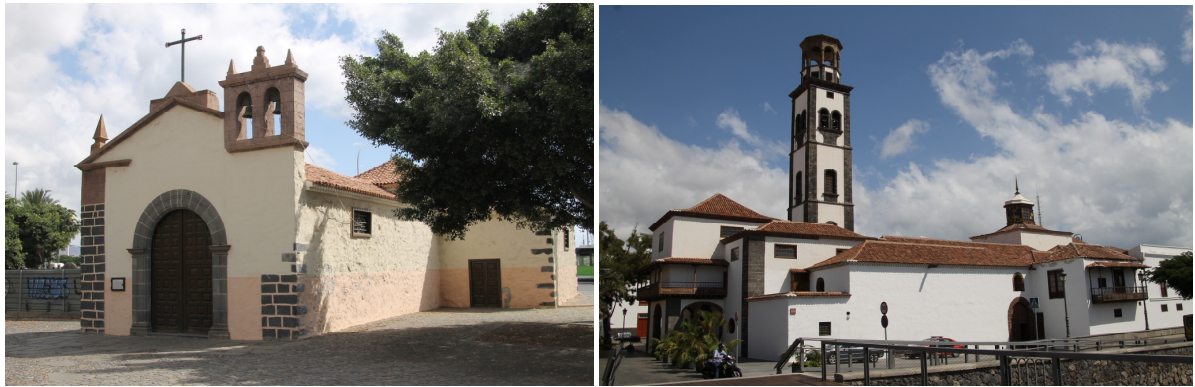

**Figure 3.** The chapel of San Telmo (left panel) and the main church of Nuestra Señora de La Concepción (right panel). The construction of both buildings dates from the 16th century and are located near the shores of Santa Cruz de Tenerife, where the main ravine, Barranco de Santos, pours its waters into the sea. Photographs by the authors.

On the other hand, half of the churches of Santa Cruz have their axes oriented to an absolute value of the declination close to 10° to 12°, similar to the orientation of La Concepción (Fig. 3), church originally founded under the advocation of the Holy Cross at the time of the conquest of the island in the very late 15th century. To the best of our knowledge, there is no evident religious festivity that could be related to this orientation (neither the Holy Cross, May 3rd, or the Immaculate Conception, December 8th, work properly), if not the celebration on August 15th corresponding to the main local feast of Our Lady of Candlemas (Nuestra Señora de La Candelaria), patroness of the island of Tenerife. Interestingly, a statue of this lady was worshiped by the Guanches in the abbreviated name of *Chaxiraxi*, mainly during the first lunation of August, in a festival called *Beñesmen*, after the time of harvest, even before the Castilian conquest of the island in 1496 (Espinosa, 1980). This would imply that the main orientation is the opposite to the canonical one, from the altar to the gate of the building, a possibility already discussed in Gangui and Belmonte (2018) for several churches in La Laguna.

However, in this particular case, the orography of the site can eventually justify this last orientation pattern (namely, churches' axes oriented within the solar range, but not precisely towards east or west), as the church of Nuestra Señora de La Concepción is parallel to the Barranco de Santos, which is a main ravine flowing from the northeast slopes of the island, where it is one of the most outstanding geographical landmarks. It had a permanent flow of water at that time, making the site a perfect location for the foundation of a new settlement that would be the natural bridgehead for the conquest of the island (La Laguna was founded ex-novo later on, once the conquest of the island was completed).

The other two churches of Santa Cruz with similar values of declination [10° to 12°], the nearly contemporary San Telmo and the church of the former (now destroyed) convent of San Francisco, could also be following this trend as does the general urban grid. However, the chapel of San Telmo (Fig. 3), which is nearly but not exactly parallel to La Concepción, does have an orientation which would have permitted the light of the setting Sun to enter across the axis of the church in dates very close to the saint's day before the Gregorian reform of the calendar (see Table 1). It is difficult to establish whether this is by chance or design but in the latter case it would be one of the few such examples found in the Canary Islands so far (there was another suspicious case in Lanzarote; Gangui et al., 2014).

## 3. CONCLUSION

From the results of this approach what stands out is a clear difference between the main patterns followed by the churches of La Laguna and the few orientations privileged by the churches of Santa Cruz de Tenerife. None of the religious buildings of the latter fall anywhere close to the main broad peak in the declination histogram we got for the churches of the former (Fig. 1). Rather, three of the churches of Santa Cruz are oriented with values of declination corresponding roughly to the position where a minimum of the La Laguna distribution is located. This result might be the tip of the iceberg bringing to light a difference in the so-far unknown initial organization of both cities, uncovering a possible deliberate plan underlying the foundation of La Laguna back at the time of the first Spanish conquerors taking possession of the island for the Crown of Castile. Indeed, a plan *not* related to a "utopian city" based on geometric principles, as proposed by popular modern traditions.

## 4. ACKNOWLEDGEMENTS

This work was partially supported by CONICET and the University of Buenos Aires, Argentina, and by the projects P/310793 'Arqueoastronomía' of the IAC, and AYA2015-66787 'Orientatio ad Sidera IV' of the Spanish MINECO.